\documentclass[aps,floats,prx,showpacs,twocolumn]{revtex4-1}

\usepackage{graphicx,subfigure}
\usepackage{amsfonts,amsmath} \usepackage{bm} \usepackage{dcolumn}
\usepackage{epsfig} \usepackage{latexsym}
\usepackage{graphicx}
\usepackage{multirow}
\usepackage{color}
\usepackage{amsmath}
\usepackage{amssymb}

\def\sp{\;,\;}
\def\nn{\nonumber}

\begin{document}

\title{Holographic model for the anomalous scalings of the cuprates}

\author{Erin Blauvelt}

\author{Sera Cremonini}

\author{Anthony Hoover}

\author{Li Li}

\author{Steven Waskie}

\affiliation{Department of Physics, Lehigh University, Bethlehem, PA, 18018 USA.}


\begin{abstract}
We examine transport in a holographic model in which the 
dynamics of the charged degrees of freedom is described by the nonlinear Dirac-Born-Infeld (DBI) 
action. Axionic scalar fields are included to break translational invariance and generate momentum dissipation in the system.
Scaling exponents are introduced by using geometries which are nonrelativistic and hyperscaling-violating in the infrared.
In the probe DBI limit 
the theory reproduces the anomalous temperature dependence of the resistivity and Hall angle of the cuprate strange metals, $\rho \sim T$ and $\cot\Theta_H \sim T^2$. 
These scaling laws would not be present without the nonlinear dynamics encoded by the DBI interactions.
We further show that because of its richness the DBI theory supports a wide spectrum of temperature scalings. 
This model provides explicit examples in which transport is controlled by different relaxation times. 
On the other hand, when only one quantity sets the temperature scale of the system, the 
Hall angle and conductivity typically exhibit the same temperature behavior.
We illustrate this point using new fully backreacted analytical dyonic black brane solutions.
\end{abstract}

\pacs{11.25.Tq, 71.27.+a, 74.72.-h}

\maketitle

\section{Introduction}
For nearly a decade holographic techniques developed within string theory have been applied to the realm of 
condensed matter physics. 
Holography has provided a novel set of analytical tools to approach many--body systems and a new 
window into the mechanisms behind strongly coupled quantum phases of matter (see e.g., Ref.~\cite{Hartnoll:2016apf} for a comprehensive review). 
The main focus of this promising research area has been on probing phase transitions and transport in models that may be in the same universality class as strongly correlated electron systems.
The latter exhibit unconventional behaviors which are believed to be tied to the complexity of their phase diagram, the presence of strong interactions and the lack of a quasiparticle description.

A prime example of such unconventional behavior is the strange metal phase of the high-temperature cuprate superconductors. 
Its anomalous features include a linear temperature dependence  for the resistivity $\rho \sim T$ \cite{Ando,Hussey,Cooper}, often believed to be associated with an underlying quantum critical point. 
Another puzzling aspect of the cuprates is the observed scaling of the Hall angle \cite{Chien,Tyler} 
$\cot\Theta_H \sim T^2$, starkly different from that of $\rho$.
These peculiar transport properties display sharp deviations from the weak coupling paradigm of Fermi liquid theory
and appear robust across different compounds.

Realizing the phenomenology of the cuprates within a holographic model 
has thus far proven to be a challenge -- in particular, reproducing the anomalous temperature dependence of both $\rho$ and $\Theta_H$ at once.
It is now understood that systems in which transport is governed by two different relaxation times should lead to different temperature behaviors for the Hall angle and  conductivity. 
For Einstein--Maxwell--Dilaton (EMD) theories this was discussed e.g., in Ref.~\cite{Blake:2014yla}.
However, even such models fail to accommodate the scaling laws of the cuprates~\cite{Amoretti:2016cad}. [Note that the examples of Ref.~\cite{Ge:2016sel} violate the null energy condition (NEC) 
while Ref.~\cite{Chen:2017gsl} used a model for which the identification of the conductivity involves a number of subtleties \cite{Cremonini:2016avj}).]
Other studies of magnetotransport based on EMD-like theories can be found 
in Refs.~\cite{Blake:2015ina,Amoretti:2015gna,Blake:2015hxa,Kim:2015wba,Zhou:2015dha,Donos:2015bxe,Lindgren:2015lia,Seo:2017yux}.

Our goal in this paper is to explore the
origin of these anomalous scalings and clarify the conditions needed to realize them. 
We work with a string-theory-motivated gravitational model~\cite{Cremonini:2017qwq} which takes into account nonlinear interactions between the charged degrees of freedom,
encoded by the Dirac-Born-Infeld (DBI) action.
Describing the low-energy dynamics of D-branes, DBI theories are nonlinear realizations of electrodynamics which are natural from a 
top--down perspective.
It is precisely the nonlinear dynamics of the gauge field sector which allows us to 
realize $\rho \sim T$ and $\cot\Theta_H \sim T^2$, and more generically a wider range of scalings. 
We emphasize that our construction is the first consistent holographic realization of the 
strange metal scalings of the resistivity and Hall angle; in particular, the latter two can be obtained simultaneously without violating the NEC.

In this model clean scaling regimes arise in a straightforward manner 
in the so-called \emph{probe limit}, in which the backreaction of the DBI interactions on the geometry can be safely neglected. 
As a consequence, gravitational solutions to the theory are of a simple form, and the resulting conductivities are easy to study analytically.
Meanwhile, the nonlinear nature of the interactions between the charged degrees of freedom is retained in this regime.
The behavior of the resistivity differs generically from that of the Hall angle because
distinct couplings control different temperature scales in the system.
Moreover, the theory admits nonrelativistic, hyperscaling-violating black brane solutions whose scaling exponents can be chosen to reproduce the cuprates. 

Realizing the same scaling laws away from this regime should not pose any conceptual challenges, but rather only technical ones. 
Finding exact analytical solutions to the theory 
in the presence of backreaction
is harder, and to do so one must rely on simplifications and 
restrictions on the parameters of the model. This can lead to a situation where only one coupling sets the temperature scale 
in the system, and controls the behavior of all the conductivities.
For these particular background solutions, then, the conductivity and Hall angle 
behave in much the same way as a function of $T$; such cases could not be used to describe the cuprates.
We illustrate this point at the end of this paper.
However, we emphasize that this is only a limitation of the analytical solutions, and numerically one can construct a much larger class of background solutions, describing systems with different 
relaxation scales.
There should be no conceptual obstacle to reproducing the phenomenology of the cuprates in these
more general settings.

In conclusion, the probe limit offers a window into the existence of clean scaling regimes, including those observed in the cuprate high-temperature superconductors.
We stress that in this class of DBI theories the cuprates' scaling laws would not be present if the 
DBI interaction was turned off; in that case the arguments developed for EMD theories would  be relevant. 
Thus, our analysis provides evidence that to capture the complexity of the phase diagram of non-Fermi liquids it may be crucial to include the nontrivial dynamics between the (charged) degrees of freedom, 
 in addition to the interplay between the various physical scales in the system.


\section{The holographic setup}

We consider a four-dimensional holographic model 
which describes gravity coupled to a neutral scalar field $\phi$, two axions $\psi^I$ and an Abelian gauge field $A_\mu$, whose dynamics is described by the DBI action,
\begin{eqnarray}\label{action}
S=&&\int d^{4}x\sqrt{-g}\left[\mathcal{R}-{1\over 2}(\partial\phi)^2-V(\phi)-\frac{Y(\phi)}{2}\sum_{I=1}^2 (\partial \psi^I)^2 \right]\nonumber\\
&&+\int d^{4}x Z_1(\phi)\Big[\sqrt{-\det(g_{\mu\nu}+Z_2(\phi)F_{\mu\nu}})\nonumber\\
&&-\sqrt{-\det (g_{\mu\nu})} \; \Big]\,,
\end{eqnarray}
where the second term in the DBI part is chosen such that in the weak flux limit $F \rightarrow 0$ one recovers the standard gauge field kinetic term.
The scalar couplings $Z_1(\phi), Z_2(\phi)$ and $Y(\phi)$ lead to nontrivial interactions between the scalar sector and the gauge field.
The axionic scalars are introduced to break translational symmetry and ensure that the system dissipates momentum and exhibits a finite DC conductivity. 
Magnetotransport in this model was studied first in Ref.~\cite{Kiritsis:2016cpm} and later in Ref.~\cite{Cremonini:2017qwq} taking into account backreaction effects.
Early work on the conductivity in probe DBI setups can be found e.g., in Refs.~\cite{Karch:2007pd,OBannon:2007cex,Hartnoll:2009ns, Lee:2010ii,Kim:2010zq,Pal:2010sx,Karch:2014mba}.

As in Ref.~\cite{Cremonini:2017qwq}, here we work with geometries of the form 
\begin{eqnarray}
\label{fullansatz}
ds^2&=&-D(r)dt^2+B(r)dr^2+C(r) (dx^2+dy^2),\;  \phi=\phi(r), \nn \\
\psi^1&=&k\, x,\; \; \psi^2=k\, y, \;\; A=A_{t}(r)\,dt+\frac{h}{2}(x dy-y dx)\,,
\end{eqnarray}
with $h$ denoting the magnitude of the magnetic field.
The linear dependence of the axions on the spatial coordinates breaks translational invariance and the strength of momentum relaxation is controlled by the parameter $k$~\cite{Andrade:2013gsa}. 
The general equations of motion are presented in Sec. I of the Supplemental Materials~\cite{sms}.
Our focus below will be on solutions which exhibit hyperscaling violation ($\theta \neq 0$) and nonrelativistic scalings ($z \neq 1$)
in the IR of the geometry, and approach anti-de Sitter (AdS) in the UV.

The DC conductivities $\sigma_{ij}$ for the theory (\ref{action}) and the background geometry (\ref{fullansatz}) were computed in Ref.~\cite{Cremonini:2017qwq}
using the horizon method developed in Refs.~\cite{Blake:2014yla,Donos:2014uba}. 
We refer the reader to  Ref.~\cite{Cremonini:2017qwq} for the analysis and report the results for $\sigma_{ij}$ in the Supplemental Materials~\cite{sms}. The key observation is that $\sigma_{ij}$ is controlled by the three scalar couplings $Z_1, Z_2, Y$ and the bulk metric component $C$, all evaluated at the horizon. 
These are generically temperature-dependent terms.
Moreover, to the extent that they are independent of each other, they in principle provide different temperature scales in the system.  
The inverse Hall angle and resistivity are then extracted by using
\begin{equation}
\cot \Theta_H=\frac{\sigma_{xx}}{\sigma_{xy}} \, , \quad \rho = \rho_{xx}=\frac{\sigma_{xx}}{\sigma_{xx}^2+\sigma_{xy}^2}\,.
\end{equation}

The conductivity associated with the DBI model is extremely rich and complex.
Also, it provides yet another example in which one does \emph{not} have the simple additive form 
$\sigma_{DC} = \sigma_{ccs} + \sigma_{diss}$. 
Indeed, the dissipative $(\sigma_{diss})$ and charge conjugation symmetric $(\sigma_{ccs})$ contributions in this model are intertwined in a nontrivial way, thanks here to the nonlinear nature of the DBI interactions.
The complexity of this DBI theory is both a challenge and an opportunity: while it is difficult to extract specific scaling properties without focusing on particularly simple sectors, one also expects to find a wide range of possible behaviors.
In particular, the transport coefficients simplify significantly in a number of limiting cases, as discussed in Ref.~\cite{Cremonini:2017qwq}. 
The one that is most relevant to us here is the probe limit.

\section{Probe DBI limit}\label{SMD}
The expressions for the conductivities of the DBI theory are much more tractable when 
the contribution to the geometry coming from the DBI sector is negligible compared to that of other matter content. 
In this case the background geometry is seeded by the scalar and axions, and the dynamics of the U(1) gauge field 
can be captured by treating it as a probe around the resulting geometry: this is the so-called probe DBI limit.
Interestingly, we find that the same expression for $\cot \Theta_H$ and $\rho_{xx}$ can be obtained from the fully backreacted case when the momentum dissipation scale $k$ dominates over the other physical scales in the system~\cite{Cremonini:2017qwq}.

In the probe DBI limit the inverse Hall angle can be seen to take the simple form
\begin{equation}
\label{strongkhall}
\cot \Theta_H=\frac{C}{h Q Z_2}\sqrt{Q^2+Z_1^2 Z_2^2(C^2+h^2 Z_2^2)}\,,
\end{equation}
and the in-plane resistivity is given by
\begin{equation}\label{rhoxx}
\rho_{xx}=\frac{C}{Z_2}\frac{\sqrt{Q^2+Z_1^2 Z_2^2(C^2+h^2 Z_2^2)}}{Q^2+C^2 Z_1^2 Z_2^2}\,,
\end{equation}
evaluated at the horizon.
Moreover, in the probe regime the charge density $Q$ and the magnetic field $h$ should be small compared to the other scales in the system (the consistency of the probe DBI approximation will be discussed in Sec. II of the Supplemental Materials~\cite{sms}).
In particular, working in the limits $Q^2 <<  Z_1^2 Z_2^2 C^2$ and $h^2 Z_2^2 << C^2$, the resistivity and the Hall angle
reduce to the very simple expressions
\begin{equation}
\label{finalapprox}
\cot \Theta_H =  \frac{C^2 Z_1}{hQ}\,, \qquad  \rho_{xx} = \frac{1}{Z_1 Z_2^2}\,, 
\end{equation}
where we have only kept leading-order terms. 
The small-$Q$ and -$h$ limits are entirely natural in the probe approximation, and will be shown below to be valid in appropriate temperature windows.

The key feature to appreciate in the expressions (\ref{finalapprox}) for $\rho_{xx}$ and $\cot\Theta_H$   is that  they generically scale differently with temperature, precisely because they are controlled by different quantities. 
The functions $C$, $Z_1$ and $Z_2$ \emph{provide different temperature scales in the system}, as long as at least two of them are independent of each other.
The technical advantage of the probe limit, as we will see shortly, is that it allows us to keep the scalar couplings $Z_1$ and $Z_2$ much more arbitrary than would be possible when working 
with specific backreacted solutions. This will give us more freedom to choose the scalings we are after.

In order to obtain the cuprates' scalings $\rho_{xx} \sim T$ and $\cot \Theta_H  \sim T^2$
from Eq.~(\ref{finalapprox}), one then needs to have a system for which
\begin{equation}
\label{cond}
\frac{C}{Z_2}=\frac{T^{3/2}}{\ell_0^{1/2}}  \quad \text{and} \quad Z_1 Z_2^2=\frac{z_0}{T}  \, ,
\end{equation}
where $\ell_0$ and $z_0$ are two positive constants that depend on the specific theory one is considering. 
Moreover, the small-$Q$ and -$h$ approximations we adopted to obtain Eq.~(\ref{finalapprox}) become, assuming a temperature dependence as in Eq.~(\ref{cond}), 
$ T>> \ell_0 Q^2/z_0^2$ and $T^3 >> \ell_0 h^2$.

At this stage it is convenient to introduce dimensionless expressions for the temperature and magnetic field, $ \mathbf{T}=\frac{z_0^2}{\ell_0 Q^2}T$ and $ \mathbf{h}=\frac{z_0^3}{\ell_0 Q^3} h$ 
respectively, as well as a constant $\zeta=\ell_0 Q^2/z_0^3$. When the condition (\ref{cond}) is satisfied, 
the expressions
\eqref{strongkhall} and \eqref{rhoxx} then become
\begin{equation}
\begin{split}
\rho_{xx}=\zeta \frac{\mathbf{T}^{3/2}}{1+\mathbf{T}} \sqrt{1+\mathbf{T}+\mathbf{h}^2/\mathbf{T}^2} ,\\
\cot\Theta_H=\frac{\mathbf{T}^{3/2}}{\mathbf{h}}\sqrt{1+\mathbf{T}+\mathbf{h}^2/\mathbf{T}^2}\,.
\end{split}
\end{equation}
It is clear that one obtains
\begin{equation}
\rho_{xx}=\zeta\, \mathbf{T},\qquad \cot \Theta_H =  \frac{\mathbf{T}^2}{\mathbf{h}}\,, 
\end{equation}
in the ``high-temperature'' limit $\mathbf{T}\gg 1+\mathbf{h}^2/\mathbf{T}^2$. 
Note that this condition is given in terms of $\mathbf{T}$, defined by using the particular scale $\ell_0 Q^2/z_0^2$ that characterizes the theory one is considering. Thus, this is not necessarily a high--$T$ limit, and it would indeed describe low temperatures 
provided that such a scale is sufficiently higher than the temperature the experiment is probing.

So far our analysis was based on the assumption that condition (\ref{cond}) could be satisfied.
We are now ready to show how it can be realized explicitly.
To proceed further we need to extract the temperature dependence of $C(r),Z_1(\phi)$ and $Z_2(\phi)$. 
Thus, we need to focus on a particular background solution and specify a choice of couplings.
In order to allow for the freedom to have scaling exponents, we are interested in geometries that are nonrelativistic 
and hyperscaling-violating in the IR, and approach AdS in the UV. 

As was shown in Ref.~\cite{Kiritsis:2016cpm}, 
when the dilaton couplings $V$ and $Y$ are approximated by exponentials in the IR,
$V(\phi)\sim -V_0 \,e^{\eta\, \phi}$ and $Y(\phi)\sim e^{\alpha\,\phi}$,
the geometry in the probe limit is of the simple hyperscaling-violating form
\begin{eqnarray}\label{axionfinitT}
ds^2&=&r^\theta\left(-f(r)\frac{dt^2}{r^{2z}}+\frac{L^2 dr^2}{r^2 f(r)}+\frac{dx^2+dy^2}{r^2}\right)\,,  \\
\phi&=&\kappa\,\ln(r),\quad \psi^1=k\,x,\quad \psi_2=k\,y\,,
\end{eqnarray}
with
\begin{eqnarray}\label{solutionhsv}
&& f(r)=1-\left(\frac{r}{r_h}\right)^{2+z-\theta}, \;  z=\frac{\alpha^2-\eta^2+1}{\alpha(\alpha-\eta)}, \; \theta=\frac{2\eta}{\alpha}\,,\nonumber\\
&&\kappa=-\frac{2}{\alpha}\, , \; L^2=\frac{(z+2-\theta)(2z-\theta)}{V_0 }\sp k^2 =\frac{2V_0(z-1)}{2z-\theta}. \nn
\end{eqnarray}
Recall that in this limit the gauge field is a probe around this background solution, and its expression can be obtained by solving
the U(1) equation of motion.
Finally, from the form of the blackening function we read off  $T \sim r_h^{-z}$ and 
\begin{equation}\label{ctotemp}
C(r_h) = r_h^{\theta-2}  \Rightarrow C(T) \sim T^{\frac{2-\theta}{z}} \, ,
\end{equation}
which is also the temperature scaling of the entropy density, $s\sim T^{\frac{2-\theta}{z}}$.

To have a well-defined geometry and a resolvable singularity one should take into account Gubser's criterion~\cite{Gubser:2000nd,Charmousis:2010zz} as well as the NEC, 
which restricts the range of $\{z, \theta\}$ appearing in Eq.~\eqref{axionfinitT}. 
Depending on the location of the IR, these restrictions yield 
 \begin{eqnarray}
\label{gubser}
&& \text{IR} \; r\rightarrow \infty : \, [1<z\leqslant 2, \theta<2z-2],\; [z>2, \theta<2]\, , \nn \\
&& \text{IR} \; r\rightarrow 0: \; \; [z \leqslant 0, \theta>2],\; [0<z<1, \theta>z+2]\,.   
\end{eqnarray}

When the backreaction of the DBI action on the geometry is taken into account, 
exact $\{z,\theta\}$ solutions to our model can be found only for particular choices of scalar couplings $Z_1(\phi)$ and $Z_2(\phi)$ (typically single exponentials). 
In the probe limit where the backreaction of the DBI sector can be neglected, there is a certain amount of freedom to choose the couplings $Z_1,Z_2$.
For simplicity---and to eventually make contact with the fully backreacted case---we take them to be 
$Z_1 \sim e^{\gamma\phi}$ and $Z_2 \sim e^{\delta \phi}$, where $\gamma,\delta$ are free parameters.  This ensures that they yield single powers of temperature when evaluated at the horizon.
Indeed, combining this with the expression for the scalar field needed to support the scaling solutions,
$ \phi = -\frac{2}{\alpha} \ln (r)$, yields 
\begin{equation}\label{zscaling}
Z_1 \sim T^{\frac{2\gamma}{z\alpha}} \quad \text{and} \quad Z_2 \sim T^{\frac{2\delta}{z\alpha}} \, .
\end{equation}
Thus, for arbitrary couplings $\gamma,\delta$ one has
\begin{equation}
\frac{C}{Z_2} \sim T^{\frac{2-\theta}{z}-\frac{2\delta}{z\alpha}}  \quad \text{and} \quad Z_1 Z_2^2 \sim T^{\frac{2\gamma + 4\delta}{z\alpha}}  \, ,
\end{equation}
and in turn
\begin{equation}
\label{genericDBIscalings}
\rho_{xx} \sim T^{-\frac{2}{z} \left( \frac{\gamma}{\alpha} + 2 \frac{\delta}{\alpha} \right)} \, ,\quad 
\cot \Theta_H \sim \frac{1}{h Q}  T^{\frac{2}{z} \left(2-\theta +  \frac{\gamma}{\alpha} \right)} \, , 
\end{equation}
for the general scaling of the resistivity and Hall angle in the probe DBI limit.
The condition required to realize the cuprates' scalings then becomes
\begin{equation}\label{condition}
\frac{\gamma}{\alpha} = z+\theta-2   \quad \text{and} \quad  \frac{\delta}{\alpha} = 1 - \frac{\theta}{2}   - \frac{3}{4} z\,.
\end{equation}
With this particular choice of Lagrangian parameters one obtains the celebrated cuprate behavior 
\begin{equation}
\label{cupratescalings}
\rho_{xx} \sim T \, , \quad \quad \cot \Theta_H \sim T^2 \, . 
\end{equation}
The validity of the probe DBI description is discussed in Sec. II of the Supplemental Materials~\cite{sms}.
We stress that there is a wide range of values of $z$ and $\theta$ (or equivalently of the theory parameters $\gamma$ and $\delta$) which satisfies all constraints and can be used to realize these 
two scaling laws. However, one still needs to identify a selection mechanism to explain why these scalings are robust and universal in the cuprates.

It is interesting to note that the $z=4/3, \theta=0$ case singled out by the purely field-theoretic analysis of Ref.~\cite{Hartnoll:2015sea} corresponds here to having $\delta=0$, or equivalently a constant $Z_2$ (and $\gamma^2 = 4/3$).
Thus, this corresponds to a minimal form of the Lagrangian, in which only the overall scalar coupling in the DBI term $Z_1 (\phi) $ is turned on. 
This case is reminiscent of the standard dilaton coupling to the 
DBI action $\sim e^{-\Phi}$ in string theory.
An interesting question is whether one could obtain the couplings needed to realize the cuprates within a top--down string theory construction. Indeed, with a UV completed theory all parameters would be entirely fixed. Note that the scaling laws~\eqref{cupratescalings} would not be present if one turned off the 
DBI interaction. Our results provide further compelling evidence for the importance of nonlinear interactions among the charge carriers for describing strange metals, as observed in other holographic models (see e.g., Ref.~\cite{Hartnoll:2009ns}).

\section{The general backreacted case}\label{Backreacted}
As we have just seen, in the probe regime these DBI models 
admit the scaling laws (\ref{cupratescalings}) observed in the cuprates,
and more generally cases in which $\rho_{xx} $ and $\cot \Theta_H $ scale differently with temperature, as in Eq.~(\ref{genericDBIscalings}). 
We expect to find the same behavior even when one moves away from the probe limit and takes into account the full backreaction of the DBI interactions on the geometry. 
However, finding exact analytical solutions that are fully backreacted is technically more challenging, and one does not expect them to be of the simple form of Eq.~(\ref{solutionhsv}), especially in the presence of a magnetic field. 
Exact analytical solutions to the DBI theory are rare, and rely on making simplifying assumptions on theory parameters.
One of the potential consequences then, is that they can lead to cases for which 
 $\sigma_{ij}$ is controlled by a single temperature-dependent quantity: a single scale. 
 In such instances one does not expect to have a 
 clean separation between the behavior of the resistivity $\rho_{xx} $ and the Hall angle $\cot \Theta_H $. 
 Indeed, the two should have a similar $T$ structure. In this section we illustrate precisely this point with an analytical example.

Exact nonrelativistic, hyperscaling-violating solutions to the full DBI theory (\ref{action}) were put forth in Ref.~\cite{Cremonini:2017qwq}.
In the presence of a background magnetic field $h\neq 0$, the scaling geometries of Ref.~\cite{Cremonini:2017qwq}
had a fixed value  of the hyperscaling-violating parameter, $\theta=4$.
Analytical solutions with arbitrary $\theta$ were also expected to exist, and to provide a more fruitful avenue to modeling possible scaling regimes.
Indeed, when the Lagrangian parameters are such that $$ \gamma = -2\delta,  \quad \eta = \alpha - \delta\, , $$ 
we have identified another class of dyonic black branes of the form (\ref{axionfinitT}), but with a blackening function given schematically by 
\begin{equation}
f(r) = 1 -(1+c_0\, r_h^{4-\theta}) \left(\frac{r}{r_h}\right)^{2 + z - \theta} + c_0\,r^{4 - \theta} \, ,
\end{equation}
where $c_0$ is a constant that depends on theory parameters, and all the remaining details of the solution are given in Sec. I of the Supplemental Materials~\cite{sms}.
We note that for these solutions the momentum dissipation parameter is not free, but is determined in terms of $h, Q$ and theory parameters.
The main feature that distinguishes this solution from that in Eq.~(\ref{solutionhsv}) is the complexity of the blackening function.
As a result, the temperature of these black branes is related to the horizon radius in a rather nontrivial way,
\begin{equation}
\label{Tstructure}
T \sim   r_h^{-z}+  \frac{c_0  (z-2)}{(2+z-\theta)}  \, r_h^{4-z-\theta}  \, ,
\end{equation}
which in turn gives a much wider range of possible temperature dependence for the entropy density than the one (\ref{ctotemp}) found in the probe limit.
The general expression (\ref{Tstructure}) is quite cumbersome, making it difficult to identify the existence of scaling regimes. 
However, in appropriate regions of parameter space only one of the two terms in Eq.~(\ref{Tstructure}) dominates,
so that one can assume a clean scaling of the form $T \sim r_h^p$
for some parameter $p$.

These exact solutions are quite constrained (they require specific relationships between theory parameters), and in particular have the property that 
the metric component $C$ and the couplings $Z_1$ and $Z_2$ are all related to each other, 
\begin{equation}
C(r) = Z_2(r) = r^{\theta-2} \, , \quad  Z_1 \sim r^{4-2\theta} = C^{-2} \, ,
\end{equation}
implying for example that the combination $Z_1 Z_2 C $ is simply a constant.
As a consequence, evaluating the conductivities on the background solutions above, we find that the temperature dependence is controlled entirely by one single quantity: the combination $CY$. Thus, this quantity sets the \emph{only} temperature scale available in the system (for early discussions of different time scales in holographic transport coefficients, see e.g., Refs.~\cite{Donos:2014uba,Gouteraux:2014hca}).
Inspecting the expressions for the conductivities, we see that the resistivity and Hall angle have the schematic form
\begin{eqnarray}
\label{CondASol}
\rho_{xx} &=& \frac{a_1 CY + a_2 (CY)^2 + a_3 (CY)^3 + a_4 (CY)^4 }{a_5 + a_6 CY + a_7 (CY)^2 + a_8 (CY)^3 + a_9 (CY)^4} \, , \nn \\
\cot\Theta_H &=& \frac{b_1 CY + b_2 (CY)^2  }{b_3+ b_4 CY + b_5 (CY)^2 } \, ,
\end{eqnarray}
where the $a_i, b_i$ are T-independent terms which depend on $h, k, Q$. 
The expressions for the coefficients are quite complicated, but all share a similar structure. 
In particular, the coefficients of $\rho_{xx}$ and $\cot\Theta_H$ in front of each power of $CY$ are generically similar to each other (for instance, the pairs $a_2$ and $b_2$, or $a_6$ and $b_4$).
What this implies is that, without \emph{severe fine-tuning} of the parameters $z_1, h$ and $Q$, one cannot generically decouple the temperature behavior of $\rho_{xx}$ from that of $\cot\Theta_H$.
The reason for this is that, unlike in the probe DBI case, the same quantity $CY$ is responsible for giving rise to all T dependence 
in this particular system.
In closing, we note that 
by fine-tuning parameters so that some of these coefficients can be made to vanish, one can indeed force $\rho_{xx}$ and $\cot\Theta_H$ to have a different scaling in terms of $CY$ (and potentially obtain the cuprates' scalings).
However, this would only hold in a very limited temperature region, and require unnatural choices of theory parameters.
This procedure would give at best a very undesirable---highly fine-tuned---realization of the scalings of the cuprates.

\acknowledgments{We would like to thank Matteo Baggioli and Blaise Gouteraux for valuable conversations.
The work of E.B., S.C., A.H. and S.W. is supported in part by the National Science Foundation grant PHY-1620169.}


\onecolumngrid
\newpage

\begin{center}
\textbf{\large Supplemental Materials}
\end{center}
\setcounter{equation}{0}
\setcounter{figure}{0}
\setcounter{table}{0}
\setcounter{page}{1}
\makeatletter
\renewcommand{\theequation}{A\arabic{equation}}
\renewcommand{\thefigure}{A\arabic{figure}}
\renewcommand{\bibnumfmt}[1]{[A#1]}
\renewcommand{\citenumfont}[1]{A#1}

\section{Equations of motion, conductivity and dyonic solutions}
\label{app:eoms}

The equations of motion associated with the action in the main text
\begin{equation}
S=\int d^{4}x\sqrt{-g}\left[\mathcal{R}-{1\over 2}(\partial\phi)^2-V(\phi)-\frac{Y(\phi)}{2}\sum_{I=1}^2 (\partial \psi^I)^2 \right]+\int d^{4}x Z_1(\phi)\left[\sqrt{-\det(g_{\mu\nu}+Z_2(\phi)F_{\mu\nu}})-\sqrt{-\det (g_{\mu\nu})} \; \right],
\end{equation}
take the form
\begin{equation}\label{eomphi}
\begin{split}
\nabla_\mu\nabla^\mu\phi-V'(\phi)-\frac{Y'(\phi) }{2}\sum_{I=1}^2 (\partial \psi^I)^2-Z_1'(\phi)
\left[\sqrt{\frac{-\det (g+Z_2(\phi) F)}{-\det g}}-1\right]  \\
+\frac{Z_1(\phi) Z_2'(\phi)}{2}\sqrt{\frac{-\det (g+Z_2(\phi) F)}{-\det g}} (g+Z_2(\phi)F)^{-1[\mu\nu]}
F_{\mu\nu}=0\,,
\end{split}
\end{equation}
\begin{equation}\label{eomA}
\nabla_\mu \left[Z_1(\phi)Z_2(\phi)\sqrt{\frac{-\det (g+Z_2(\phi) F)}{-\det g}}(g+Z_2(\phi)F)^{-1[\mu\nu]}\right]=0\,,
\end{equation}
\begin{equation}\label{eompsi}
\nabla_\mu\left(Y(\phi)\nabla^\mu \psi^I\right)=0\, ,
\end{equation}
%
%
\begin{equation}\label{eomg}
\begin{split}
\mathcal{R}_{\mu\nu}-{1\over 2}\mathcal{R}g_{\mu\nu}=\frac{1}{2}\left(\partial_\mu\phi\partial_\nu\phi-\frac{1}{2}g_{\mu\nu}(\partial\phi)^2\right)
+\frac{Y(\phi)}{2}\sum_{I=1}^2\left(\partial_\mu\psi^I\partial_\nu\psi^I-\frac{1}{2}g_{\mu\nu}(\partial\psi^I)^2\right)\\
-\frac{1}{2}g_{\mu\nu}V(\phi) +T^{DBI}_{\mu\nu}\,,
\end{split}
\end{equation}
with the DBI stress energy tensor given by
\begin{equation}\label{tensor}
T^{DBI}_{\mu\nu}=
-\frac{Z_1(\phi)}{2}\sqrt{\frac{-\det (g+Z_2(\phi) F)}{-\det g}}g_{\mu\alpha}(g+Z_2(\phi) F)^{-1(\alpha\beta)}g_{\beta\nu}+\frac{Z_1(\phi)}{2}g_{\mu\nu}\,.
\end{equation} 
Here $(g+Z_2(\phi)F)^{-1\mu\nu}$ is the inverse of $(g+Z_2(\phi)F)_{\mu\nu}$, with the subscript $^{(\;)}$ denoting the symmetric part (and $^{[\;]}$ the antisymmetric part).
The current in the dual field theory, evaluated at the boundary, reads
\begin{equation}\label{current}
\begin{split}
J^\mu&=\sqrt{-\gamma} \, n_\nu \,  Z_1(\phi)Z_2(\phi)\sqrt{\frac{-\det (g+Z_2(\phi) F)}{-\det g}}  \left(g+Z_2(\phi)F \right)^{-1[\nu\mu]} \Big{|}_\partial\,,\\
&=Z_1(\phi)Z_2(\phi)\sqrt{-\det (g+Z_2(\phi) F)} \left(g+Z_2(\phi)F\right)^{-1[r\mu]} \Big{|}_\partial\,.
\end{split}
\end{equation}
The quantities $\gamma$ and $n^\mu$ in this expression are, respectively, the induced metric and outward pointing normal vector at the asymptotically AdS boundary. Here we have used $r$ to denote the holographic radial direction.

Assuming homogeneity and isotropy, we have taken the bulk metric and the matter fields to be the generic form,
\begin{eqnarray}\label{smfullansatz}
\begin{split}
ds^2=-D(r)dt^2+B(r)dr^2+C(r) (dx^2+dy^2),\quad \phi=\phi(r),\\
\psi^1=k\, x,\quad\psi^2=k\, y,\quad A=A_{t}(r)\,dt+\frac{h}{2}(x dy-y dx)\,,
\end{split}
\end{eqnarray}
with $h$ denoting the magnitude of the magnetic field.
Substituting this ansatz into~\eqref{eomphi}-\eqref{eomg}, we obtain the following equations:
\begin{equation}\label{eqphi}
\begin{split}
\frac{1}{\sqrt{BD}C}\left(\sqrt{\frac{D}{B}}C\phi'\right)'+\frac{\Omega}{C\sqrt{BD}}\frac{Z_2'(\phi)}{ Z_2(\phi)}\left((C^2+2 h^2 Z_2(\phi)^2)A_t'^2-h^2 BD\right)\\
-Z_1'(\phi)\left(\frac{Z_1(\phi)Z_2(\phi)^2}{\Omega C\sqrt{BD}}-1\right)-\frac{k^2 }{C}Y'(\phi)-V'(\phi)=0\,,
\end{split}
\end{equation}
\begin{equation}\label{eqmetric1}
\frac{D'C'}{DC}+\frac{1}{2}\frac{C'^2}{C^2}-{1\over 2}\phi'^2+B Z_1(\phi)\left(\frac{\Omega C\sqrt{BD}}{Z_1(\phi)Z_2(\phi)^2}-1\right)+\frac{\Omega B\sqrt{BD}h^2}{C}+\frac{k^2 B}{C}Y(\phi)+B V(\phi)=0\,,
\end{equation}
\begin{equation}\label{eqmetric2}
\frac{2C''}{C}-\left(\frac{B'}{B}+\frac{C'}{C}+\frac{D'}{D}\right)\frac{C'}{C}+\phi'^2=0\,,
\end{equation}
\begin{equation}\label{eqmetric3}
\frac{2D''}{D}-\frac{2C''}{C}-\left(\frac{B'}{B}-\frac{C'}{C}+\frac{D'}{D}\right)\frac{D'}{D}+\frac{B'C'}{BC}-2\Omega\sqrt{BD}\left(\frac{CA_t'^2}{D}+\frac{B h^2}{C}\right)-\frac{2 k^2 B}{C}Y(\phi)=0\,,
\end{equation}
\begin{equation}\label{eqA}
\Omega (C^2+h^2 Z_2(\phi)^2) A_t'=Q\,,
\end{equation}
where $Q$ is the charge density and we have introduced the function
\begin{equation}\label{omega}
\Omega(r)= \frac{Z_1(\phi) Z_2(\phi)^2}{\sqrt{(C^2+h^2 Z_2(\phi)^2)(BD-Z_2(\phi)^2 A_t'^2)}} \, .
\end{equation}

The conductivity matrix for the background geometry~\eqref{smfullansatz} is given by
\begin{equation}\label{DC}
\begin{split}
\sigma_{xx}&=\sigma_{yy} = \frac{k^2 C Y \left[\Omega(h^2\Omega+k^2 Y)(C^2+h^2 Z_2^2)^2+C^2Q^2\right]}{(h^2\Omega+k^2 Y)^2 (C^2+h^2 Z_2^2)^2+h^2 C^2 Q^2}\, ,\\
\sigma_{xy}&=-\sigma_{yx}
=\frac{hQ[(h^2\Omega+k^2 Y)^2(h^2 Z_2^4+2 C^2Z_2^2)+(h^2\Omega+k^2 Y)C^4 \Omega+C^2Q^2-C^2k^2Y(C^2 \Omega+k^2 Y Z_2^2)]}{(h^2\Omega+k^2 Y)^2 (C^2+h^2 Z_2^2)^2+h^2 C^2 Q^2} \,,
\end{split}
\end{equation}
where it is understood that all functions should be evaluated at the horizon $r_h$. We have also adopted a metric parametrization for which $B=1/D$ as in Ref.~\cite{Cremonini:2017li}, and therefore~\eqref{omega} can be written as
\begin{equation}
\Omega(r) = \frac{Z_2}{C^2 + h^2 Z_2^2} \sqrt{Q^2 + Z_1^2 Z_2^2 (C^2 + h^2 Z_2^2)} \, .
\end{equation}
We point out that the conductivity as a function of the temperature $T$ does not depend on the metric parametrization. 

The fully backreacted dyonic geometry discussed in the main text is given by
\begin{equation}
\begin{split}
ds^2 &= r^\theta\left(-f(r)\frac{dt^2}{r^{2z}}+\frac{L^2 dr^2}{r^2 f(r)}+\frac{dx^2+dy^2}{r^2}\right)\, ,\\
\phi&=\kappa\,\ln(r)\,,\; \psi^1=k\,x \, ,\;\psi_2=k\,y\, ,  \; A = A_t(r) dt + \frac{h}{2} (x dy - y dx) \, ,
\end{split}
\end{equation}
with
\begin{align}
\label{AHsolution}
& f(r) = 1 - \left(\frac{r}{r_h}\right)^{2 + z - \theta} + \frac{2(z-1)(2 + z - \theta)(h^2z_1^2 + Q^2)}{k^2(z - 2)(2z + \theta - 6)\sqrt{(1 + h^2)z_1^2 + Q^2}}r^{4 - \theta}\left[1 -\left(\frac{r}{r_h}\right)^{z - 2}\right] \, ,  \nn \\
& \kappa = -\frac{2}{\alpha} \, , \quad z = \frac{1+\alpha^2-\eta^2}{\alpha(\alpha-\eta)} \, , \quad
\theta = 2 \, \frac{\eta}{\alpha} \, , \quad 
k^2 = \frac{V_0}{1- \frac{(\eta-\alpha)^2-1}{ (\eta^2-\alpha \eta -1)} \sqrt{(1+h^2)+\frac{Q^2}{z_1^2}} }  \, , \\
&   A_t'(r) = \frac{L Q}{z_1} \frac{r^{1 - z}}{\sqrt{(1 + h^2) + \frac{\rho^2}{z_1^2}}} \, ,   
\quad L^2 = \frac{2(z-1)(2 + z - \theta)}{k^2}  \, .\nn
\end{align}
Here we have used $Z_1=z_1 e^{\gamma\phi}$, $Z_2=e^{\delta\phi}$, $Y=e^{\alpha\phi}$ and $V(\phi)=-V_0 e^{\eta\phi}$ with $z_1$ and $V_0$ two positive constants. The temperature of these black brane solutions is given by
\begin{equation}
T =  \frac{1}{4\pi }  \sqrt{\frac{2+z-\theta}{2(z-1)}} \left[k \,  r_h^{-z}+ \frac{2(z-1)}{k(\theta+2z-6)}  \frac{h^2 z_1^2 + Q^2}{\sqrt{1+h^2 z_1^2+ Q^2}}  \, r_h^{4-z-\theta} \right] \, .
\end{equation}

\section{The validity of the probe DBI approximation}\label{app:probe}

In this section we discuss the validity of the probe DBI limit. 
In particular, one should demand that the contribution of the DBI terms should
be subleading to the gravity background, and the geometry is seeded by the scalar $\phi$ and axions $\psi^I$. More precisely, we require the stress tensor of the DBI action to be much smaller than the Einstein tensor.

Using an analysis similar to that in Ref.~\cite{Kiritsis:2016}, one obtains the constraint
\begin{equation}\label{constraint}
Z_1^2 Z_2^2 r_h^{2\theta} L^2\ll Z_2 r_h^2\sqrt{Q^2+Z_1^2 Z_2^2(C^2+h^2 Z_2^2)}\ll Z_2^2/L^2\,,
\end{equation}
evaluated at the horizon $r_h$. Recall that to reproduce the cuprate strange metal scalings, 
we need to fix the temperature dependence of $(C, Z_1, Z_2)$. 
Using the dimensionless temperature and magnetic field defined in the main text, we then obtain
\begin{equation}
N_1\,\mathbf{T}^{\frac{z-4}{z}}  \ll\mathbf{T}^{-\frac{2\theta+3 z}{2 z}}\sqrt{1+\mathbf{T}+\mathbf{h}^2/\mathbf{T}^2}\ll N_2\, \mathbf{T}^{\frac{4-2\theta-3 z}{z}}\,,
\end{equation}
where $N_1$ and $N_2$ are constants that are independent of $\mathbf{T}$ and $\mathbf{h}$. 

Since we are interested in the high temperature limit  $\mathbf{T}\gg 1+\mathbf{h}^2/\mathbf{T}^2$, we find
\begin{equation}\label{probe}
\mathbf{T}^{\frac{4-\theta-2 z}{z}}\gg 1\quad \Rightarrow \quad {\frac{4-\theta-2 z}{z}}>0\,.
\end{equation}
We point out that we neglected the last term in the DBI stress energy tensor~\eqref{tensor} when deriving the constraint above. 
This is consistent with the probe approximation and indeed, as one can check, $V\gg Z_1$ in the regime we are interested in.
To resolve the naked singularity which is present in the hyperscaling-violating ground state, we already considered Gubser's criterion and the null energy condition, which impose non-trivial constraints on $(z,\theta)$. 
The validity of the probe approximation then imposes further constraint~\eqref{probe} on the two scaling exponents.
It is important to take into account all such constraints, in order to have a consistent parameter space for the
holographic theory. Indeed, we find that there exists a large range of parameter space for $(z,\theta)$ satisfying all the above constraints. Finally, the magnetic field can not be too strong, with the upper bound given by $\mathbf{h} \ll \mathbf{T}^{3/2}$.

\end{document}